\documentstyle[preprint,aps,epsfig]{revtex}

\begin{document}

\title{Coulomb blockade in a quantum wire with long-range interactions}
\author{H{\'e}l{\`e}ne Maurey and Thierry Giamarchi}
\address{Laboratoire de Physique des Solides, Universit{\'e} Paris--Sud,
                   B{\^a}t. 510, 91405 Orsay, France\cite{junk}}
\maketitle

\begin{abstract}
We study the transport through two impurities or ``barriers''
in a one-dimensional quantum wire, taking into account the long-range
$\frac1r$ Coulomb interactions. We compute the temperature-dependent
conductance $G(T)$ of this system.
Long-range forces lead to a dramatic increase of weak barrier potentials
with decreasing temperature, even in the
``resonant'' case. The system thus always reaches a ``strong barrier''
regime in which only charge is pinned, contrary to the standard LL case.
$G(T)$ vanishes faster than any power as $T$ goes to zero.
In particular, resonant tunneling is suppressed at zero temperature.
\end{abstract}


Hope for experimental characterization of Luttinger
Liquids(LL) \cite{haldane_bosonisation} has been renewed
by the fabrication of nanostructures such as quantum wires
\cite{wires_gen}.
 A LL is expected for short-range Coulomb interactions. However, in an isolated
 quantum wire, interactions can be long-range,
 leading to a Wigner Crystal(WC) with dominant $4k_F$
 fluctuations and quasi-long-range order \cite{schulz_wigner_1d}.
For long wires containing many impurities, one expects
universal $T^2$ dependence of the conductivity for the WC \cite{maurey_qwire}
versus interaction-dependent power-law for the LL \cite{many_imps}.
Attention focussed recently on short wires with a few
impurities. For one impurity and repulsive interactions,
conductance vanishes at zero $T$,
as a power-law in the LL \cite{kane_qwires_tunnel,furusaki_1imp},
faster than a power-law in the WC
\cite{glazman_single_impurity,fabrizio_furusaki}.
With two impurities, only LL was studied
\cite{kane_qwires_tunnel,furusaki_2imp}.
In the strong impurity regime, Coulomb Blockade and resonant tunneling phenomena
occur. On resonance, perfect transmission is expected at $T=0$ for moderately repulsive
interactions.

We examine here the double-barrier in presence of long-range
interactions, a situation relevant to realistic quantum wires.
We show that in this case, contrarily to the LL, the physics is
drastically different depending on whether one applies
directly a very strong impurity potential $V \gg E_F$, or one starts
from an initially weak potential ($V \ll E_F$) which is renormalized
to large values as $T$ decreases.
We show that the latter case leads, for the WC or LL with strongly
repulsive interactions,
to blockade of charge, spin being free to
flow, whereas in the former both charge and spin are locked.
We focus here on two initially ``weak'' impurities, which is the
relevant case for most experimental situations (due to intrinsic
disorder or artificial constrictions), the ``strong'' impurity case
corresponding to tunnel junctions in the wire.
We show that only charge degree of freedom
plays a role in transport and that a WC exhibits Coulomb Blockade and
charge resonances and compute microscopically the
charging energy of the island formed between the two barriers.
We draw the parallel with the phenomenological
charging energy term usually added to explain Coulomb Blockade in standard
mesoscopic systems \cite{devoret_averin}.
Unlike in a LL
this charging energy is dominated by electrostatic effects.
Both for weak and strong barriers, using respectively perturbation theory
and an instanton method, we obtain
the temperature-dependent conductance on and away from resonance.
The conductance always
vanishes faster than any power with decreasing $T$, even in the resonant case.
At $T=0$ resonant tunneling is suppressed.
These effects could provide an
experimental signature of the importance of long-range Coulomb interactions
and the existence of the one-dimensional WC.

We consider a narrow wire of interacting electrons, of
width $d \sim \lambda_F$ and
of length $L \gg d$ so that the system is regarded as one-dimensional.
The hamiltonian of the pure system is
%
\begin{equation} \label{hamtotal}
H  = \sum_{\nu=\rho,\sigma}{u_{\nu} \over 2\pi }\int _{-\frac L2}^{\frac L2}
 dx \lbrack K_\nu(\pi \Pi_{\nu})^2
  + {1 \over  K_{\nu}} (\partial _x \Phi_{\nu} )^2 \rbrack
+ \frac1{\pi ^2} \int _{-\frac L2}^{\frac L2}
  \int _{-\frac L2}^{\frac L2} dxdx'
V(x-x') (\partial _x\Phi_{\rho}(x)) (\partial _{x}\Phi_{\rho}(x'))
\end{equation}
%
We have taken $\hbar =1$, $\nu=\rho,\sigma$ indicate the charge and spin
degrees of freedom. $\Pi_{\nu}$ is the conjugate momentum of
$\Phi_{\nu}$, $K_{\rho}$ a number giving the strength
of the short-range part of Coulomb forces, $u_{\rho}$ is the
renormalized Fermi velocity due to the same interactions.
We assume spin-isotropic case, $K_{\sigma}=1$.
$V(x)=\frac{e^2}{\kappa\sqrt{x^2+d^2}}$ is the  long range part of the
Coulomb interaction ($\kappa$ is the dielectric constant). (\ref{hamtotal}) describes
a one dimensional WC, dominated by $4 k_F$
charge fluctuations.
Indeed $4 k_F$ charge correlation functions
are always the slowest decaying ones, with much
slower decay than power laws \cite{schulz_wigner_1d}. $2 k_F$ ones are
still power-laws due to the spin part.
Screened long-range interactions correspond to
$V=0$ in (\ref{hamtotal}), which in that case describes a LL:
charge and spin correlations decay as power laws, with interaction-dependent
exponents. $2 k_F$ fluctuations dominate
for $K_{\rho}>\frac13$ and $4 k_F$ ones for $K_{\rho}<\frac13$ .

We model two impurities by delta functions $V_1\delta(x+\frac a2)$ and
$V_2\delta(x-\frac a2)$. As long as the renormalized barriers are weak
($\ll E_F$), we trivially keep only coupling
to the slowest decaying part of the density
\begin{equation} \label{imp}
H_{\text{imp}} \simeq V_{1,4k_F} \cos\bigl(\sqrt{8}\Phi_{\rho}(-\frac
a2)+2k_Fa\bigr)+ V_{2,4k_F} \cos\bigl(\sqrt{8}\Phi_{\rho}(\frac a2)-2k_Fa\bigr)
\end{equation}
where $V_{i,4k_F}=V_i\rho(4k_F)$.
Furthermore, even if such weak barriers flow to strong coupling,
working with (\ref{imp}) only is still valid. In particular we needn't
take into account the
terms $V_{2k_F}\cos(\sqrt{2}\Phi_{\rho}(\pm\frac a2))
\cos(\sqrt{2}\Phi_{\sigma}(\pm\frac a2))$. For simplicity, let us show
this for a single impurity case, a full derivation being given in
\cite{maurey_strong_imp}. Since
$V_{4k_F}$ is the most relevant
term in the potential \cite{fabrizio_furusaki},  $V_{4k_F}/E_F$
becomes of order one while
$V_{2k_F}/E_F$ is still very small and can be treated perturbatively.
The minimization of
$V_{4k_F}\cos(\sqrt{8}\Phi_{\rho})$ imposes that
$V_{2k_F}\cos(\sqrt{2}\Phi_{\rho})\cos(\sqrt{2}\Phi_{\sigma})$ is zero.
More precisely, one can show using the instanton method introduced in the
following that effective $V_{2k_F}$ in this regime decreases exponentially as
$T \to 0$ \cite{maurey_strong_imp}.
Taking {\it initially} a strong impurity would be very different:
one then has to minimize the sum of {\it all}
harmonics in the potential.
Minima are connected by $(\Phi_{\rho},\Phi_{\sigma}) \to
(\Phi_{\rho} \pm \frac{\pi}{\sqrt{2}},
\Phi_{\sigma} \pm \frac{\pi}{\sqrt{2}})$: thus
{\it here both charge and spin are locked}, giving
different $G(T)$ \cite{fabrizio_furusaki}.
For the WC a weak link
is not the asymptotic limit of a weak impurity problem.
An analog distinction was made in
the context of the X-ray edge singularity \cite{oreg_fes}.

Depending on temperature, the two barriers in (\ref{imp}) act in
conjugation or as independent
scatterers, since
finite $T$ introduces a thermal length above which
correlations are destroyed. Using the dispersion relation, to link energy
and positional quantities, one obtains for the
WC, the thermal length
$L_T \sim \frac{u\sqrt{\alpha_c}}{T} \ln^{1/2}\frac{T_d}T$
where $\alpha_c=\frac{4Ke^2}{\pi u\kappa}$ (from now on we drop the
subscript $\rho$ since only charge excitations will matter).
$T_d=\frac{u\sqrt{\alpha_c}}d$ is a cut-off temperature
($L_T \sim \frac{u}{T}$ for a LL).
When $L_T<a$ the two barriers are uncorrelated and
transport is very similar as for a single barrier,
whereas if $L_T>a$ the barriers
exhibit new behavior due to mutual coupling.
In this regime,
and for very weak barriers the WC is almost not distorted but chooses a
relative position with respect to the barriers minimizing potential
energy. $\Phi$ is therefore pinned, except for
 $\frac{4k_Fa}{2\pi} = N+\frac12$ ($N$ integer)
{\it and} $V_{1,4k_F}=V_{2,4k_F} \equiv V_0$ for which there is no
preferred position: the WC slides freely at $T=0$. Such ``resonance''
occurs only
for symmetric barriers, situation considered in the following
(weak asymmetric barriers can be assimilated to the off resonance case
\cite{maurey_strong_imp}). Using
perturbation in powers of $V_0$ one computes the corrections $\delta G$
to the conductance $G_0$ of the pure case\cite{cond_pur2}.

\begin{figure}
  \centerline{\epsfig{file=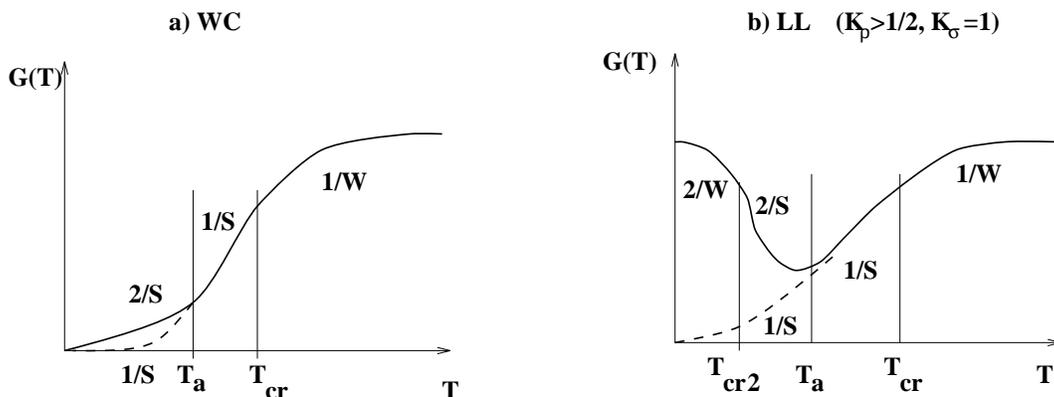,angle=-90,width=14cm}}
   \caption{Temperature dependence of the conductance through a
symmetric double-barrier
    for a WC (a) and a LL with $K_{\rho}>\frac12$ (b), starting from
weak barriers at a given $T$
    (right of the figure).
   We show here the situation $T_a<T_{cr}$ when the first crossover is to the
   strong-single-barrier regime. In the LL case there is a second crossover to
   weak coupling for $T<T_{\text{cr2}}$.
   Dashed (full) lines give off (on-)resonance cases.
   (Charge resonance for WC, spin (Kondo) resonance for LL).
In ``1 or 2/W or S'',
   1 stands for one impurity or two impurity case off resonance, 2 for two impurities on resonance,
   W and S mean weak or strong coupling regimes. Expressions for $G(T)$ are
   (\ref{1W}), (\ref{2W}), (\ref{simult}), (\ref{conduct}) for
the WC and in Ref.[7]
   for the LL.}

      \label{compar}
\end{figure}

At $T > T_a\sim \frac{u\sqrt{\alpha_c}}{a}\ln^{1/2}\frac ad$
($T_a\sim \frac ua$ for the LL) only correlations involving one
barrier contribute, one is in the single-barrier regime:
\begin{equation}\label{1W}
\delta G(T) \sim -\gamma g^2T^{-2}\ln^{-1/2}\frac{T_d}T
e^{-4\nu\ln^{1/2}\frac{T_d}T}
\end{equation}
where $\gamma$ is a constant of order $e^2/h$ and
$\nu=\sqrt{\frac{\pi u\kappa K}{e^2}}$.The temperature dependence is identical
to the one found for many weak impurities at high temperature
\cite{maurey_qwire,giamarchi_one_to_many}.  A WC is more strongly pinned
by a single barrier than a LL for which divergence of $\delta G$ at low $T$
is slower ($T^{K_{\rho}-1}$ for electrons with spin
\cite{kane_qwires_tunnel,furusaki_1imp}).
When $\delta G$ becomes of the order of $G_0$,
there is crossover to a strong-coupling regime. Two cases
occur depending on $T_a$:
either each barrier first flows to strong coupling
at $T_{cr} > T_a$
while the system remains in the single barrier regime, or there is first a
crossover to the double barrier regime at $T_a$. In
such ``weak-double-barrier'' regime $T < T_a$ one has to
consider correlations between the two barriers.
Out of resonance, $\delta G$ grows as in the single-barrier case (\ref{1W}) times a
factor $(1+\cos4k_Fa)$ up to crossover to strong coupling.
Close enough to resonance
 the $8 k_F$ component of $H_{\text{imp}}$ gives the main contribution
\begin{equation}\label{2W}
G(T)_{\text{on-res}} \sim G_0 -\gamma (1+\cos8k_Fa)g^2T^{-2}\ln^{-1/2}\frac{T_d}T
e^{-16\nu\ln^{1/2}\frac{T_d}T}
\end{equation}
$G$ decreases though slower than off-resonance.
Contrary to a LL with moderate repulsive interactions $1/2<K_{\rho}<1$
for which $\delta G$ vanishes till perfect resonant transmission at $T=0$
(see fig.\ref{compar} and \cite{kane_qwires_tunnel,furusaki_1imp}),
the WC always crosses over to strong coupling. In this regime the
barriers coincide with minima of the $4k_F$ part of the charge density,
by compression or dilatation of the WC.
Smallest distortion then defines the ground
state on the island. It is unique and corresponds to $N$ electrons,
except on resonance ($\frac{4k_Fa}{2\pi}=N+\frac12$, as for weak
symmetric barriers), with two degenerate ground states for
$N$ and $N+1$ electrons. Again we treat symmetric barriers.
Transport through the double-barrier is determined by the
evolution of $\Phi$ at the locations
of the barriers and one can integrate all other modes to obtain
the action for $\Phi(\pm a/2)$. When $T > T_a$,
 $\Phi(-\frac a2)$ and $\Phi(\frac a2)$
decouple in the effective action and we recover the ``single-barrier''
case:
$G(T) \sim t^2T^{-2}\ln^{-1}\frac{T_d}T
e^{-\frac43 \frac1{\nu}\ln^{3/2}\frac{T_d}T}$
where $t$ is the amplitude of tunneling through one barrier
(a power $T^{-1}$ was found in \cite{fabrizio_furusaki},
which is correct for {\it initially strong} ($V \gg E_F$) impurity as said above,
due to a large $V_{2k_F}\gg E_F$ term).
  Such decrease is faster than any power (for a LL
$G(T) \sim T^{\frac1{K_{\rho}}-1}$). For $T < T_a$ the two barriers
are coupled and it is simpler to
introduce  $\tilde{\Phi} \equiv \big(\Phi(-\frac a2)-\Phi(\frac
a2)\big)$ describing the
state on the island and
$\overline{\Phi} \equiv \frac12\big(\Phi(-\frac a2)+\Phi(\frac
a2)\big)$ describing the island seen from the leads.
The effective action reads:
\begin{eqnarray}\label{tf}
{\cal S}_{\text{eff}}&=&   \frac12(\frac{m_\Lambda}2)\int d\tau
\dot{\tilde{\Phi}}(\tau)^2
+\frac12(2m_\Lambda)\int d\tau \dot{\overline{\Phi}}(\tau)^2
+ 2g\int d\tau \cos(\sqrt{8}\overline{\Phi}(\tau))
\cos(\sqrt{2}\tilde{\Phi}(\tau)+2k_Fa)
+{\cal S}_{\tilde{\Phi}}+{\cal S}_{\overline{\Phi}}+{\cal S}_c\\
{\cal S}_c&=&\frac1{2{\cal C}}\int d\tau
\bigl( \frac{e\sqrt{8}}{2\pi}\tilde{\Phi}(\tau)\bigr)^2\\
{\cal S}_{\tilde{\Phi}}&=& \frac1{4\pi^2\nu}\ln^2\frac ad
\int \int d\tau d\tau'
\ln^{-3/2}\Biggl(\frac{|\tau - \tau'|}{\tau_c}\Biggr)
 \frac{\bigl(\tilde{\Phi}(\tau)-\tilde{\Phi}(\tau')\bigr)^2}{(\tau - \tau')^2}\\
 \label{last}
{\cal S}_{\overline{\Phi}}&=&
\frac1{\pi^2\nu}\int \int d\tau d\tau'
\ln^{1/2}\Biggl(\frac{|\tau - \tau'|}{\tau_c}\Biggr)
 \frac{\bigl(\overline{\Phi}(\tau)-\overline{\Phi}(\tau')\bigr)^2}{(\tau - \tau')^2}
\end{eqnarray}
$\Lambda$ is a momentum cut-off, $\tau_c^{-1}=u\Lambda$,
$m_\Lambda=\frac{4\sqrt{\alpha_c}}{\pi^2\Lambda u K}$,
${\cal C}=\frac a2 \ln^{-1} \frac ad$.
For large $g$,
$(\sqrt{8}\overline{\Phi},\sqrt{2}\tilde{\Phi})$ is constrained
to minima of the potential
$(\pi + n_0\pi, \pi+ m_0\pi -2k_fa)$, with $n_0+m_0$ {\it odd}.
Transport is possible only by quantum tunneling of electrons through the
potential barriers, corresponding to instantons of
the phases $(\sqrt{8}\overline{\Phi},\sqrt{2}\tilde{\Phi})$,
$(\pm \pi,\pm \pi)$ if an electron crosses {\it one} barrier,
$(\pm 2\pi,0)$ if two electrons tunnel simultaneously, one through
each barrier. This can be described
 in a WKB approximation \cite{schmid_instanton}, or instanton method:
the equation of motion for $\Phi$
obtained from (\ref{tf}) is solved on each barrier independently.
The general solution for
$\Phi(\pm \frac a2,\tau)$ is a linear combination of instantons
and anti-instantons.

The first part (\ref{tf}) of $S_{\text{eff}}$ describes
a ``gas of non-interacting''
instantons. But ${\cal S}_{\overline{\Phi}}$, ${\cal S}_{\tilde{\Phi}}$
and ${\cal S}_c$, due to electron-electron interactions, give correlations
between tunneling events. ${\cal S}_{\overline{\Phi}}$
and ${\cal S}_{\tilde{\Phi}}$ provide instanton-instanton repulsion and
instanton-anti-instanton attraction.
${\cal S}_c$ limits the number of electrons added to the island and
is of the form $\int Q^2/2{\cal C}$.
$Q=\frac{e\sqrt{8}}{2\pi}\tilde{\Phi}$ is variation of charge
in a segment of length $a$ of the perfect WC. $Q^2/2{\cal C}$
is the electrostatic energy of the island and
${\cal C}$ its capacitance. Such
charging energy is responsible for Coulomb blockade.
It is due to Coulomb repulsion cost of creating an excess
or lack of charge and to the elastic cost of the corresponding
distortion. In the WC,
Coulomb repulsion cost dominates and the blockade is genuinely of
electrostatic nature. In
the spinless LL, most of the ``charging'' energy comes from kinetic energy
of the electrons, i.e. the quantization of levels on the island
\cite{kane_qwires_tunnel}, and short range interactions only provide
small enhancement.
Moreover for the WC both island and leads
contribute (in equal part) to the charging energy, whereas in the LL
only the inner island is involved. Indeed due to long-range
interactions the island charge creates an image one on the leads
which contributes to the charging energy.

After substitution of the instanton form of $\tilde{\Phi}$
and $\overline{\Phi}$ in
(\ref{tf})-(\ref{last}) the partition function can be computed:
\begin{eqnarray}
\label{partition}
Z&=&\sum_{\{n_1,n_2\}} \sum_{\{q_i\},\{r_i\}} t^{2n_T}
\int_0^{\beta}d\tau_{2n}...\int_0^{\tau_2}d\tau_1
 e^{\frac 4{3\nu}\Bigl(\sum_{i<j}q_iq_j
\ln^{3/2} \frac {|\tau_i - \tau_j|}{\tilde{d}}
-(3\ln^2 \frac ad) r_ir_j \ln^{-1/2} \frac {|\tau_i - \tau_j|}{\tilde{d}}\Bigr)-{\cal S}_c}
\end{eqnarray}
$t=e^{-s_0}$ is the
amplitude of tunneling of an electron through {\it one} barrier,
$s_0=\sqrt{8m_{\Lambda}g}$ the action of a single instanton.
$Z$ is summed over all possible paths containing $n_1$(resp. $n_2$)
events (instantons or anti-instantons) through the left (resp. right)
barrier. $n_T=n_1+n_2$.
At time $\tau_i$, $q_i$ electron(s) cross the double-barrier from left
to right and $2r_i$ are
added to the island.
(\ref{partition}) is
perturbatively expanded in terms of the number of tunneling events.
Writing the current as
$I=\frac {e\sqrt{8}}{2\pi}\dot{\overline{\Phi}}$, this allows to compute the
conductance.
Two regimes occur
depending on whether Coulomb blockade is determinant or not. If the cost
of an additional electron is higher than thermal
energy, the number of electrons on the island is
locked to $N$ and the main transport process is
{\it simultaneous tunneling} in which one electron is transferred
directly from one lead to the other ({$q_1=\pm 1$, $r_1=0$}), giving
\begin{equation} \label{simult}
G_{\text{off res.}}(T) \sim t^4T^{-2}\ln^{-1}\frac{T_d}Te^{- \frac
4{3\nu}\ln^{3/2}\frac{T_d}T}
\end{equation}
Although temperature dependence is the same as in the one barrier
regime, tunneling is strongly reduced since the amplitude goes
as $t^4$ instead of $t^2$.
Oppositely, when Coulomb Blockade term is negligible compared
to thermal fluctuations, transport mechanism is {\it two-step tunneling}
where an electron crosses one barrier after the other
($(q_1,r_1)=(+\frac12,\frac12)$, and $(q_2,r_2)=(+\frac12,-\frac12)$, i.e.
charge on the island hops between $N$ and $N+1$
giving
\begin{equation}
\label{conduct}
G_{\text{on res.}}(T) \simeq t^2T^{-2}\ln^{-1}\frac{T_d}T
e^{- \frac1{3\nu}\ln^{3/2}\frac{T_d}T}
\end{equation}
$G(T)$ vanishes faster than any power, though
much slower than off resonance and with a factor $t^2$ instead of $t^4$,
owing to two-step tunneling.

The physics of the WC is thus very different from that of a LL with spin
in which both charge and spin are locked, and despite the spin degrees
of freedom closer to a spinless LL. In the LL with spin,
lowest-order resonances are spin instead of charge ones, analog to Kondo
resonances, allowing perfect transmission of both charge
and spin at $T=0$ if $K_{\rho}>1/2$ (fig.\ref{compar}b)
\cite{kane_qwires_tunnel}. The conditions for resonance are also
different: half-integer number of electrons
between the barriers for the charge resonance in the WC, odd number
for the spin resonance in the LL.
Conductance oscillations are expected to vanish in a WC
when lowering $T$, whereas they should deepen in a LL with not too strong
interactions.
These experimentally testable effects could therefore allow to probe
the strength and nature of interactions in a one-dimensional wire.


\end{document}